\documentclass [prb,superscriptaddress, showpacs, preprint]{revtex4-1}
\usepackage{graphicx}
\usepackage{dcolumn}

\newcommand{\CACR}{$\alpha$-CaCr$_2$O$_4$}
\newcommand{\half}{\frac{1}{2}}
\newcommand{\quarter}{\frac{1}{4}}
\begin{document}

\title{Helical magnetic state in the distorted triangular lattice of \CACR\ }

\author{L.C. Chapon}
\author{P. Manuel}
\affiliation{ISIS facility, STFC Rutherford Appleton Laboratory, Chilton, Didcot, Oxfordshire, OX11-0QX,United Kingdom}
\author{F. Damay}
\affiliation{Laboratoire Leon Brillouin, CEA-CNRS, UMR 12, CEA-Saclay, 91191 Gif-sur-Yvette Cedex, France}
\author{P. Toledano}
\affiliation{Laboratory of Physics of Complex Systems, University of Picardie, 33 rue Saint-Leu, 80000 Amiens, France}
\author{V. Hardy}
\affiliation{Laboratoire CRISMAT, ENSICAEN, UMR 6508 CNRS, 6 Boulevard du Marechal Juin, 14050 Caen Cedex, France}
\author{C. Martin}
\affiliation{Laboratoire CRISMAT, ENSICAEN, UMR 6508 CNRS, 6 Boulevard du Marechal Juin, 14050 Caen Cedex, France}
\date{\today}

\begin{abstract}
The magnetic properties of the high temperature $\alpha$ form of the CaCr$_2$O$_4$ compound have been investigated for the first time by magnetic susceptibility, specific heat and powder neutron diffraction. The system undergoes a unique magnetic phase transition at 43K to a long range order incommensurate helical phase with magnetic propagation vector $\textbf{k}=(0,0.3317(2),0)$. The magnetic model proposed from neutron diffraction data shows that the plane of rotation of the spins is perpendicular to the wave-vector, and that the magnetic modulation is consistent with two modes belonging to distinct irreducible representations of the group. The magnetic point group 2221' is not compatible with ferroelectricity unlike the CuCrO$_2$ delafossite [\textit{Kimura et al., Phys. Rev. B, 78 140401 (2008)}] but predicts the existence of quadratic magnetoelectric effects, discussed based on a Landau analysis.
\end{abstract}


\maketitle
\section{Introduction}
\indent In recent years, spin-driven ferroelectricity given rise to \emph{giant} magnetoelectric effects has been found in a variety of frustrated magnets, in particular in systems that develop long-wavelength magnetic modulations such as cycloidal or helicoidal states, incommensurate or not with the crystalline lattice. Most of the attention was initially focussed on cycloidal magnets such as TbMnO$_3$ \cite{ISI:000231310900063} for which the inverse Dzyaloshinskii-Moriya \cite{PhysRevB.73.094434} and spin-current model \cite{PhysRevLett.95.057205} predict the onset of a spontaneous polarization when the direction of the modulation ($\textbf{k}$-vector) is perpendicular to the spin-rotation axis ($\textbf{e}$) \cite{PhysRevLett.96.067601}. More recent work on incommensurate magnetic phases in triangular lattices \cite{PhysRevLett.98.267205,PhysRevB.78.140401,PhysRevB.73.220401,PhysRevB.77.052401,PhysRevB.79.214423,
PhysRevB.79.014412} has established that improper ferroelectric order of similar magnitude also appears in spirals magnets, i.e systems that possess a true magnetic chirality ($\textbf{k}$ parallel to $\textbf{e}$), however requiring another microscopic coupling mechanism such as hybridization effects proposed by Arima \cite{JPSJ.76.073702}. Those systems are quasi two-dimensional triangular lattices and display a nearly 120$^{\circ}$ magnetic arrangement in the triangular layers as expected for Heisenberg spins. Improper  ferroelectricity predicted by symmetry \cite{PhysRevB.79.014412} arguments depends only on the orientation of $\textbf{e}$ with respect to the crystal axes, often dictated by weak single-ion anisotropy terms.\\   
\indent In this paper, we report for the first time on the magnetic properties of \CACR\, whose orthorhombic crystal structure \cite{CaCr2O4Structure} is closely related to the structure of delafossites, at the exception of a small distortion of the triangular lattice. The magnetic susceptibility indicates strong antiferromagnetic correlations (Weiss temperature of -893K) and a unique transition to a long-range ordered state at T$_N$=43K. The specific heat displays a sharp and intense peak around 43K, consistent with a unique transition. Powder neutron diffraction proves the existence of long-range magnetic order below T$_N$, incommensurate with the crystal lattice with propagation vector $\textbf{k}$=(0,0.3317(2),0). The model proposed for the magnetic structure, derived from the neutron data and symmetry considerations, corresponds to a nearly 120 $^{\circ}$ \emph{out of plane} helix, with \textbf{e} perpendicular to \textbf{k}. Symmetry analysis using the complete irreducible co-representations of the wave-vector group shows that the point group symmetry of the magnetic state is 2221' and can be only stabilized through two successive second-order transitions or through a first-order transition. Whilst in the rhombohedral delafossite such out-of-plane spiral leaves a polar point group, here the spin chirality preserves all proper rotations, predicting the absence of electric polarization at T$_N$. However, we show through a Landau analysis of the magnetoelectric free energy that quadratic magnetoelectric effects are allowed by symmetry and should be observed in low magnetic fields.\\
\section{Experimental}
\indent The synthesis of polycrystalline \CACR\ followed a two-steps process. A sample of $\beta$-CaCr$_2$O$_4$  was first prepared in the shape of a rod (6 millimeters in diameter and several centimeters in length), starting from a 1:1 mixture of CaO and Cr$_2$O$_3$ heated at 1400$^{\circ}$C for 12 hours in an argon flow. This rod was then heated in argon atmosphere by using an image furnace to obtain the high temperature conditions (T$>$1700$^{\circ}$C) but without reaching the melting temperature. Magnetic susceptibility has been measured using a Superconducting Quantum Interference Device  (SQUID,Quantum Design $\textregistered$) in a magnetic field of 0.3T on warming following a zero-field cooling process. Heat capacity measurements (C(T)) were carried out in a Physical Property Measurement System (PPMS,Quantum Design $\textregistered$) using a semi-adiabatic relaxation method.\cite{Cp1} Two types of procedures were followed to record zero-field C(T) curves: (i) equally spaced data points registered upon warming and using a small temperature rise of the order of 0.4 K around the transition. The analysis is the standard one in which the so-called 2$\tau$ model \cite{Cp2} is used to fit at once both the heating and cooling branches at each point; (ii) a large single-pulse method described in \cite{Cp3} where the temperature is swept across the complete width of the transition, either warming or cooling (temperature rise of 5K). It this case, the heat capacity is derived from a point-by-point analysis of the time response along each of the two branches. In case of a first-order transition, it is known that the former method can yield spurious results,\cite{Cp4,Cp5} whereas the latter is able to account for both the latent heat and hysteretical features.\cite{Cp3} Neutron diffraction experiments were collected on the WISH diffractometer of the ISIS Facility, Rutherford Appleton Laboratory (UK). Data have been focus on five histograms from detector banks covering 32  $^{\circ}$ each, and are shown for the bank centred at 2$\theta$=90$^{\circ}$. A 0.55g sample was placed in a thin wall (30$\mu$m) vanadium cylindrical can mounted in an helium cryostat. Diffractograms have been recorded in the paramagnetic regime at 70K, at 1.5K and warming in varying temperature steps. Rietveld refinements of the neutron data have been performed with the FullProf program \cite{FullProf}.\\
\section{Results}
\begin{figure}[h!]
\includegraphics[scale=0.40,angle=-90.0]{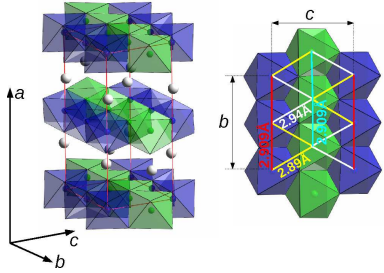}
\caption{(Color online) Left) Crystal structure of \CACR. CrO$_6$ octahedra are shown in blue and green colors for the inequivalent Cr$_1$ and Cr$_2$ sites (see text for details). The Ca$^{2+}$ ions are shown as light grey spheres. The red thin line marks the crystallographic unit-cell. Right) Projection of the CrO$_2$ layer onto the \textit{bc}-plane. Colored thick lines shows the connectivity between first-neighbor Cr ions. The corresponding interactomic distances are shown.}
\label{fig:structure}
\end{figure}
\begin{figure}
\includegraphics[scale=0.82]{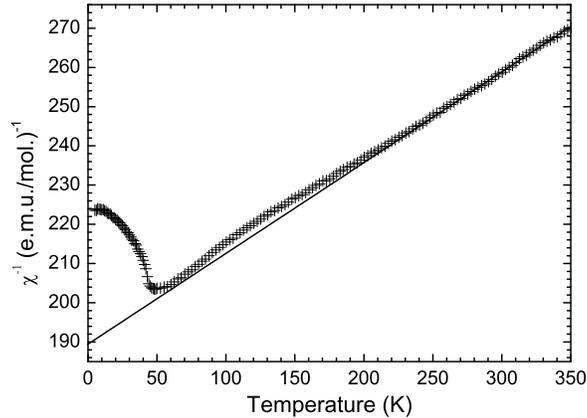}
\caption{Inverse molar magnetic susceptibility of \CACR\ as a function of temperature under a magnetic field of 0.3T. The line is a linear fit with a Curie-Weiss law. }
\label{fig:chi}
\end{figure}
\begin{figure}
\includegraphics[scale=0.40]{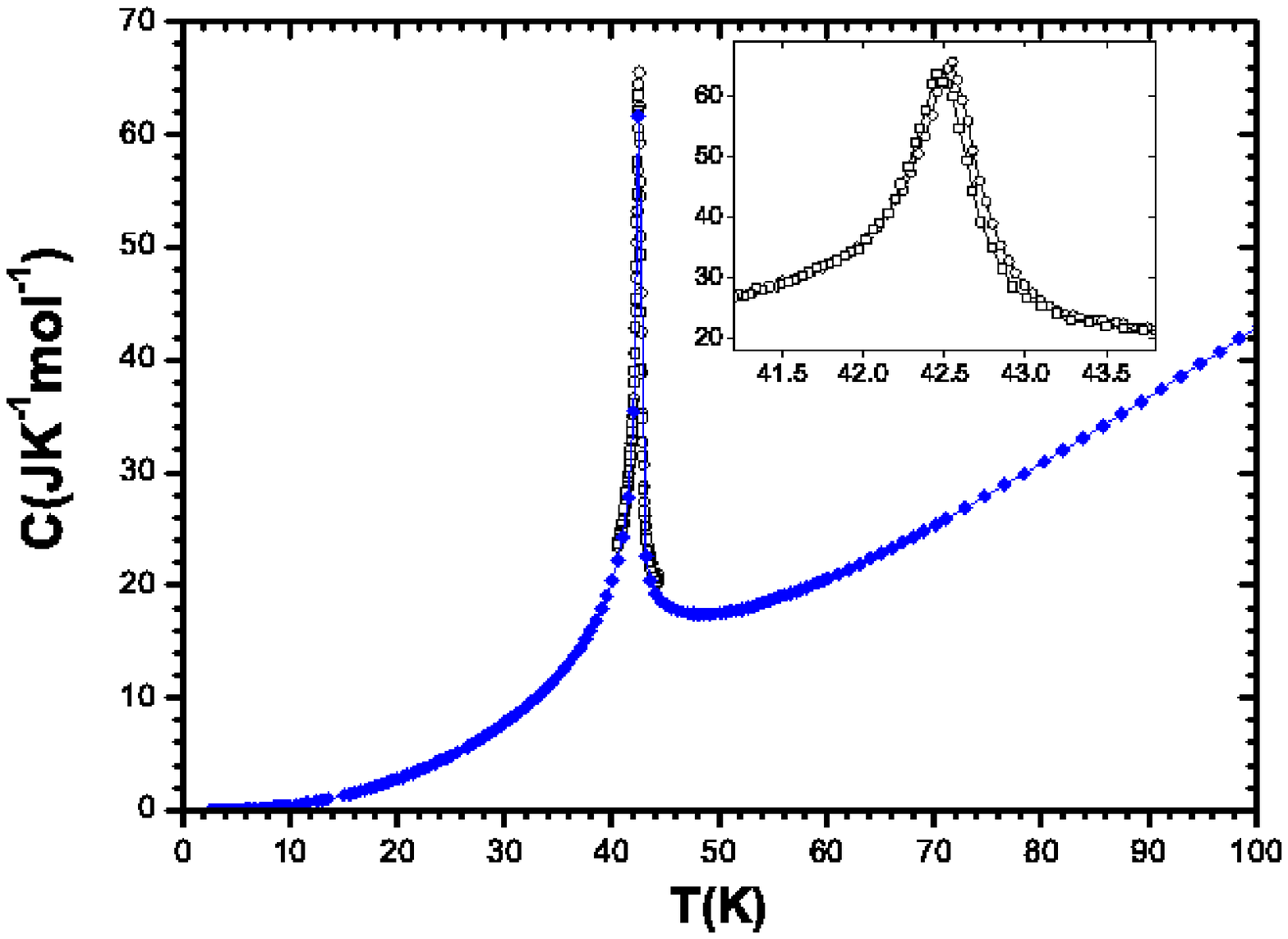}
\includegraphics[scale=0.40]{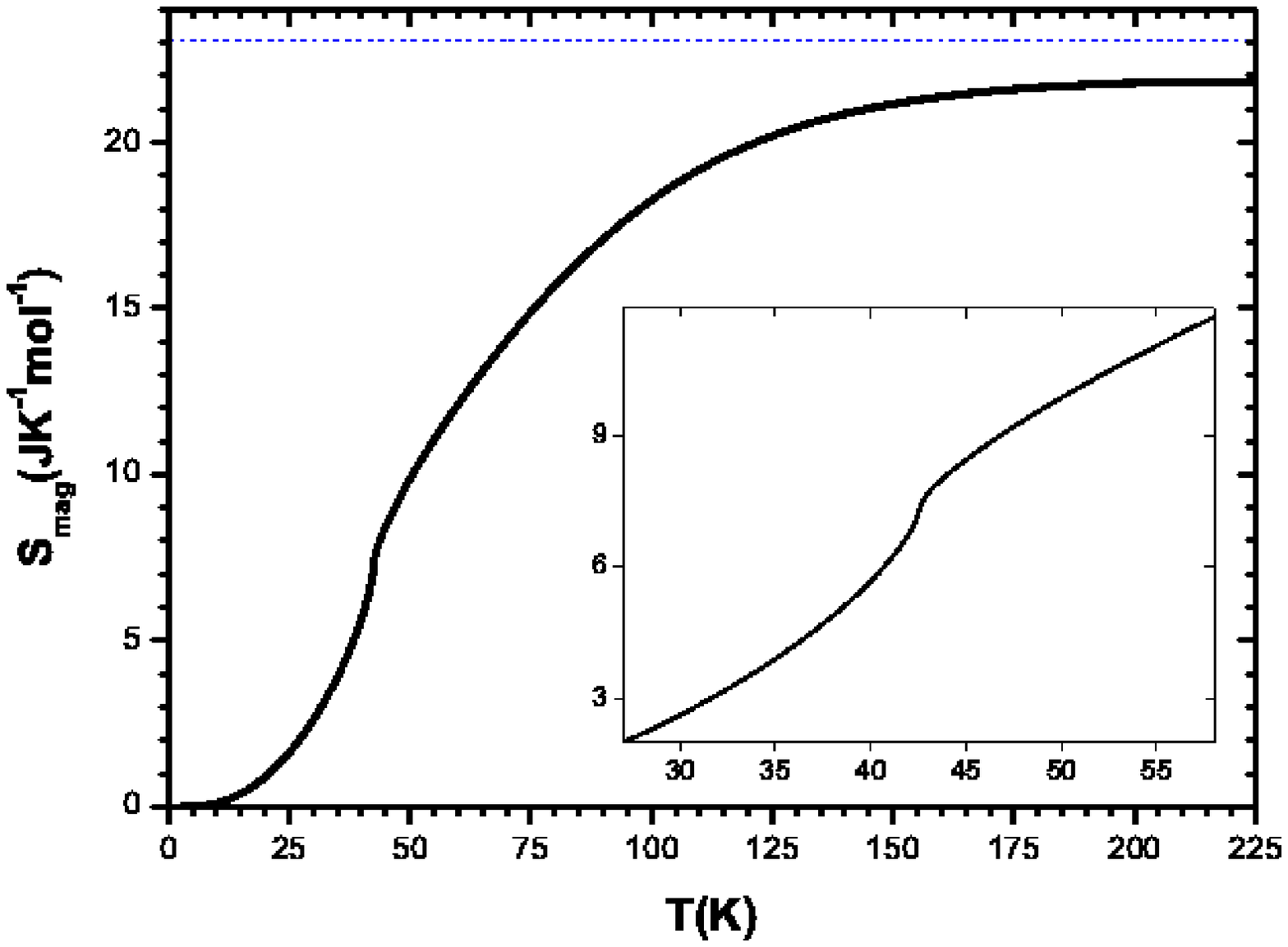}
\caption{Left: Specific heat of \CACR\ in zero magnetic field measured by the standard (filled diamonds), single pulse method (open circles) and SPM cooling (open squares) (see text for details of the various techniques). Right: Magnetic entropy obtained by subtracting the lattice contribution estimated by a Debye function to the total specific heat. The inset focus on the temperature region close to T$_N$. The dashed line indicates the theoretical limit of 2Rln(4).}
\label{fig:cp}
\end{figure}
\begin{figure}[h!]
\includegraphics[scale=0.24]{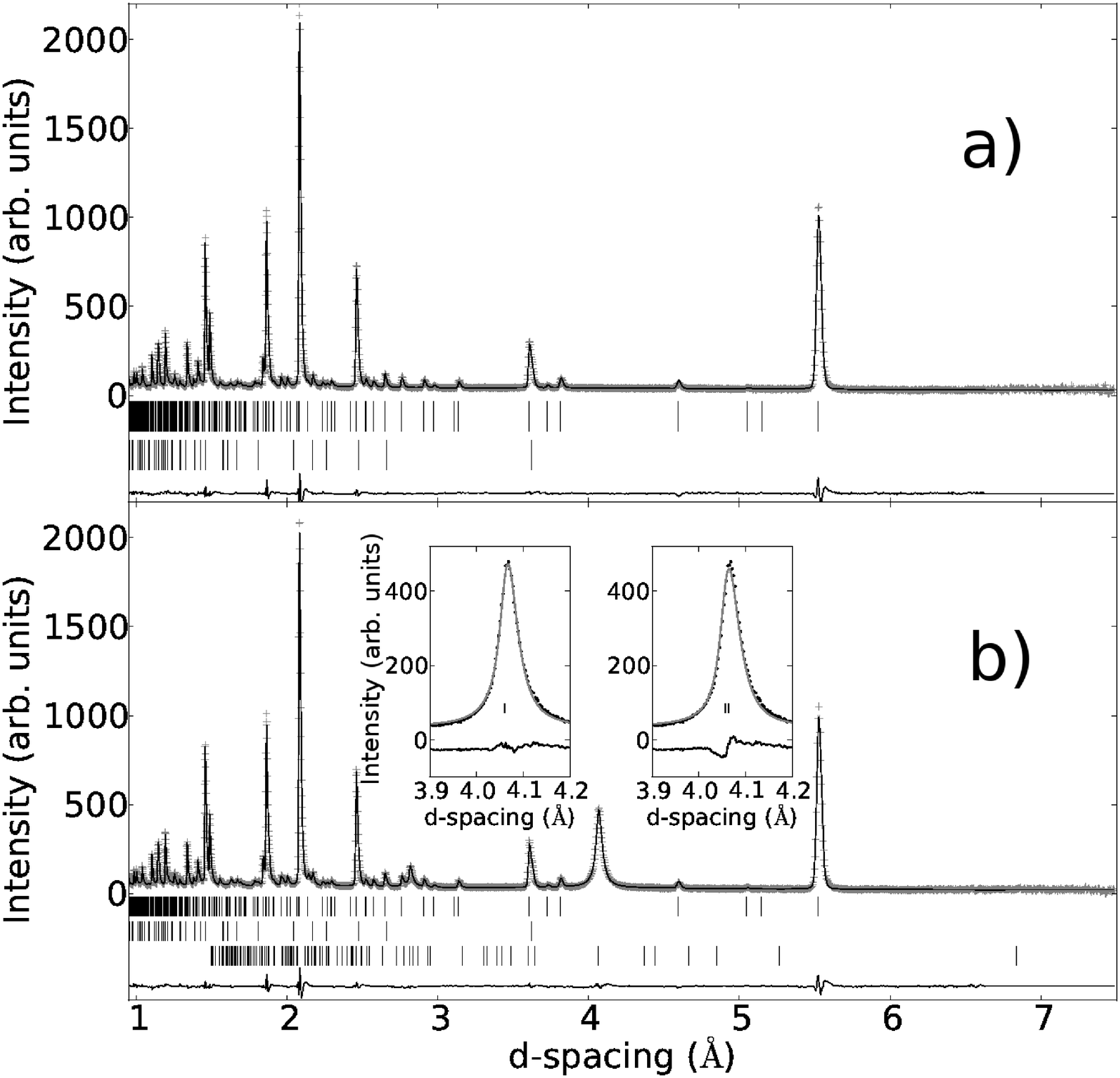}
\caption{Rietveld refinements of the neutron diffraction data collected in the paramagnetic phase at 70K (a) and in the magnetically ordered phase at 1.5K (b). The data are shown as grey points and the result of the refinements as continuous black lines. The curve shifted below represent the differences between observed and calculated patterns. In both panels, the top two rows of tick marks indicate the positions of Bragg peaks for the phases \CACR\ and the impurity Cr$_2$O$_3$. In panel b, the third row of tick marks indicated the positions of magnetic Bragg peaks with propagation vector k=(0,0.33,0). }
\label{fig:rietveld}
\end{figure}

The crystal structure of \CACR\ was first reported by Pausch et al. \cite{CaCr2O4Structure}. It crystallizes with orthorhombic symmetry, confirmed by our Electron Microscopy measurements (not shown) in the space group $Pmmn$ (Fig. \ref{fig:structure}). The structure refined from our neutron diffraction data in the paramagnetic regime at 70K is also consistent with the earlier work \cite{CaCr2O4Structure}. There are two inequivalent Cr sites, Cr$_1$ located at (0,0,0), of $\overline{1}$ symmetry and Cr$_2$ at position (0.5047(6),$\frac{1}{4}$,0.497(1)) located on a mirror plane, each site of multiplicity four. The Cr ions are six-coordinated by oxygen in a distorted octahedral configuration. The octahedra share edges to form CrO$_2$ layers in the $bc$-plane, leaving a weakly distorted triangular lattice as shown by the different interatomic distances marked in Fig. \ref{fig:structure}. The CrO$_2$ layers, separated by Ca$^{2+}$ ions, are stacked along the $a$-axis of the orthorhombic unit-cell. The in-plane magnetic interactions between Cr can be mediated either by direct exchange through the t$_{2g}$ orbitals or by Cr-O-Cr super-exchange interactions. The interactions between layers can be mediated by super-superexchange interactions through two oxygen atoms, but are expected to be much weaker due to the large  interatomic Fe-Fe distance (5.52 $\AA$ at 1.5K).\\
\indent The inverse magnetic susceptibility (Fig. \ref{fig:chi}) shows a quasi-linear regime above 250K. A linear fit of the data with a Curie-Weiss law in the temperature range 250K to 350K yields a paramagnetic moment of 4.3$\mu_B$ per Cr$^{3+}$ ion, and a Weiss temperature of -893K. The value of the paramagnetic moment is larger than the expected value of 3.87$\mu_B$ for a pure spin contribution (S=3/2) of Cr$^{3+}$. The reason for this discrepancy is likely related to the limited temperature range of the measurement, much lower than the Weiss temperature, and it seems rather unphysical to ascribe this deviation to a large orbital contribution. The strongly negative Weiss temperature indicates large antiferromagnetic (AFM) interactions. On cooling below 200K,one observe a deviation of the inverse susceptibility from the nearly linear regime followed at T$_N$=43K by a more abrupt drop indicative of long-range AFM ordering. There are no evidence of additional transitions from susceptibility measurements at different fields (not shown). The large difference between the value of the Weiss temperature and T$_N$ is inherent to the reduction of dimensionality in this pseudo-2D system as well as magnetic frustration imposed by the triangular geometry.\\                   
\indent The left panel of Fig. \ref{fig:cp} displays the specific heat, C(T), recorded from the standard and single-pulse methods (see experimental section), while the inset is an enlargement exhibiting  the shift between the warming and cooling branches derived from the latter technique. At T$_N$=43K, a very sharp peak is observed with a full width at half maximum of about 0.5 K. Using a Debye function to fit the high temperature data as an approximation of the lattice specific heat, one can derive the temperature dependence of the magnetic entropy (S$_m$) shown in right panel of Fig. \ref{fig:cp}. There is a clear kink at T$_N$ on the Sm(T) curve, but judging the nature of the transition from this feature only is difficult. The Sm(T) around the transition (Fig. \ref{fig:cp} inset) can indeed be regarded either as a broadened jump – expected for a first-order transition (FOT) in “real” materials, i.e. non ideals with the presence of defects- or as a knee –typical of a second-order transition (SOT). Two other experimental observations must be considered: first, the nearly perfect superimposition of the C(T) curves derived from the single-pulse and standard method, whilst a substantial part of the latent heat -if present- is supposed to be invisible in the latter technique.\cite{Cp3} Secondly, the temperature shift observed between the heating and cooling data in the single-pulse method is very small, and can not even be safely ascribed to a genuine hysteresis as encountered in most of the FOT’s. In fact, it appears that this hysteresis quantitatively corresponds to the expected temperature lag between the sample and the platform which contains the thermometer. At this stage, one must conclude that the transition at T$_N$ is either a SOT or a very weekly FOT. Further information might be derived from the analysis of the “critical” behavior around T$_N$, but this will require measuring C(T) for an isostructural non-magnetic compound to provide an accurate estimate of the lattice contribution more precise than a simple extrapolation of a Debye function. This work is currently in progress.\\      
\indent Below 43K, long-range magnetic order is detected by the presence of five additional Bragg peaks in the powder neutron diffraction data, the most intense close to a d-spacing of 4.5 $\AA$. The intensity of these peaks decreases with the modulus of the scattering vector (Q) in agreement with the expected Q-dependence of the magnetic form factor. To determine the periodicity of the magnetic structure, an automatic indexing procedure using a grid search \cite{FullProf} was employed but failed to give a satisfactory solution due to the presence of pseudo-trigonal symmetry. However, the magnetic peaks could all be indexed by exploring reciprocal space in the vicinity of the special k-vector expected for a 120$^{\circ}$ structure in a isotropic triangular lattice, corresponding to $\textbf{k}$=(0,$\frac{1}{3}$,0) in the \CACR\ orthorhombic unit-cell. Further analysis shows that the magnetic structure is actually incommensurate with a propagation vector $\textbf{k}$=(0,0.3317(2),0), labelled k$_8$ in Kovalev's notation \cite{Kovalev}. The incommensuration, albeit small, is genuine as indicated by the inferior refinement presented in Fig. \ref{fig:rietveld} showing an offset for the position of the magnetic peak (most noticeable in the difference curve) when the value is locked at $\textbf{k}$=(0,$\frac{1}{3}$,0). The profile of the magnetic peaks are almost purely Lorentzian and their widths are much larger than the instrumental resolution, indicating short correlations lengths. In order to account for the observed profile and derive an accurate value for the moment magnitudes, the Rietveld refinement of the magnetic phase was conducted using a phenomenological description of the peak broadening by a spherical harmonics expansion consistent with the Laue class mmm. 
\begin{table*}[h]
\begin{tabular}{c|c|c|c|c|c|c|c|c}
Irrep. & \{2$_x\vert\half$ 00 \} & \{2$_y\vert$0$\half$0\} & \{2$_z\vert\half \half$0\} & \{$\overline{1}\vert$000\}
& \{m$_{y}\vert\half$ 00\}& \{m$_{xz}\vert$0$\half$ 0\} & \{m$_{xy}\vert \half \half$ 0\} & \{1$\vert$010\} \\
\hline
$\Delta_1$ & 
$\left(\begin{array}{cc}
0 & 1 \\
1 & 0 \\
\end{array}\right)$ &
$\left(\begin{array}{cc}
\epsilon & 0 \\
0 & \epsilon^* \\
\end{array}\right)$ & 
$\left(\begin{array}{cc}
0 & \epsilon^* \\
\epsilon & 0 \\
\end{array}\right)$ &
$\left(\begin{array}{cc}
0 & 1 \\
1 & 0 \\
\end{array}\right)$ &
$\left(\begin{array}{cc}
1 & 0 \\
0 & 1 \\
\end{array}\right)$ &
$\left(\begin{array}{cc}
0 & \epsilon^* \\
\epsilon & 0 \\
\end{array}\right)$ &
$\left(\begin{array}{cc}
\epsilon & 0 \\
0 & \epsilon^* \\
\end{array}\right)$ &
$\left(\begin{array}{cc}
\epsilon^2 & 0 \\
0 & \epsilon^{*2} \\
\end{array}\right)$ \\
\hline
$\Delta_2$ & 
$\left(\begin{array}{cc}
0 & -1 \\
-1 & 0 \\
\end{array}\right)$ &
$\left(\begin{array}{cc}
\epsilon & 0 \\
0 & \epsilon^* \\
\end{array}\right)$ & 
$\left(\begin{array}{cc}
0 & -\epsilon^* \\
-\epsilon & 0 \\
\end{array}\right)$ &
$\left(\begin{array}{cc}
0 & 1 \\
1 & 0 \\
\end{array}\right)$ &
$\left(\begin{array}{cc}
-1 & 0 \\
0 & -1 \\
\end{array}\right)$ &
$\left(\begin{array}{cc}
0 & \epsilon^* \\
\epsilon & 0 \\
\end{array}\right)$ &
$\left(\begin{array}{cc}
-\epsilon & 0 \\
0 & -\epsilon^* \\
\end{array}\right)$ &
$\left(\begin{array}{cc}
\epsilon^2 & 0 \\
0 & \epsilon^{*2} \\
\end{array}\right)$ \\
\end{tabular}
\caption{Matrix representatives of the complete irreducible corepresentations $\Delta_1$ and $\Delta_2$ associated with the irreducible representations of the little group $\tau_1$ and $\tau_2$ \cite{Kovalev} (see text for details). The matrices act each on the two-dimensional spaces spans by the complex order parameters, respectively ($\eta_1$,$\eta_1^*$) and ($\eta_2$,$\eta_2^*$). The matrix for time-reversal is not shown.}
\label{Table:irrep}
\end{table*} 

\begin{figure}
\includegraphics[scale=0.37]{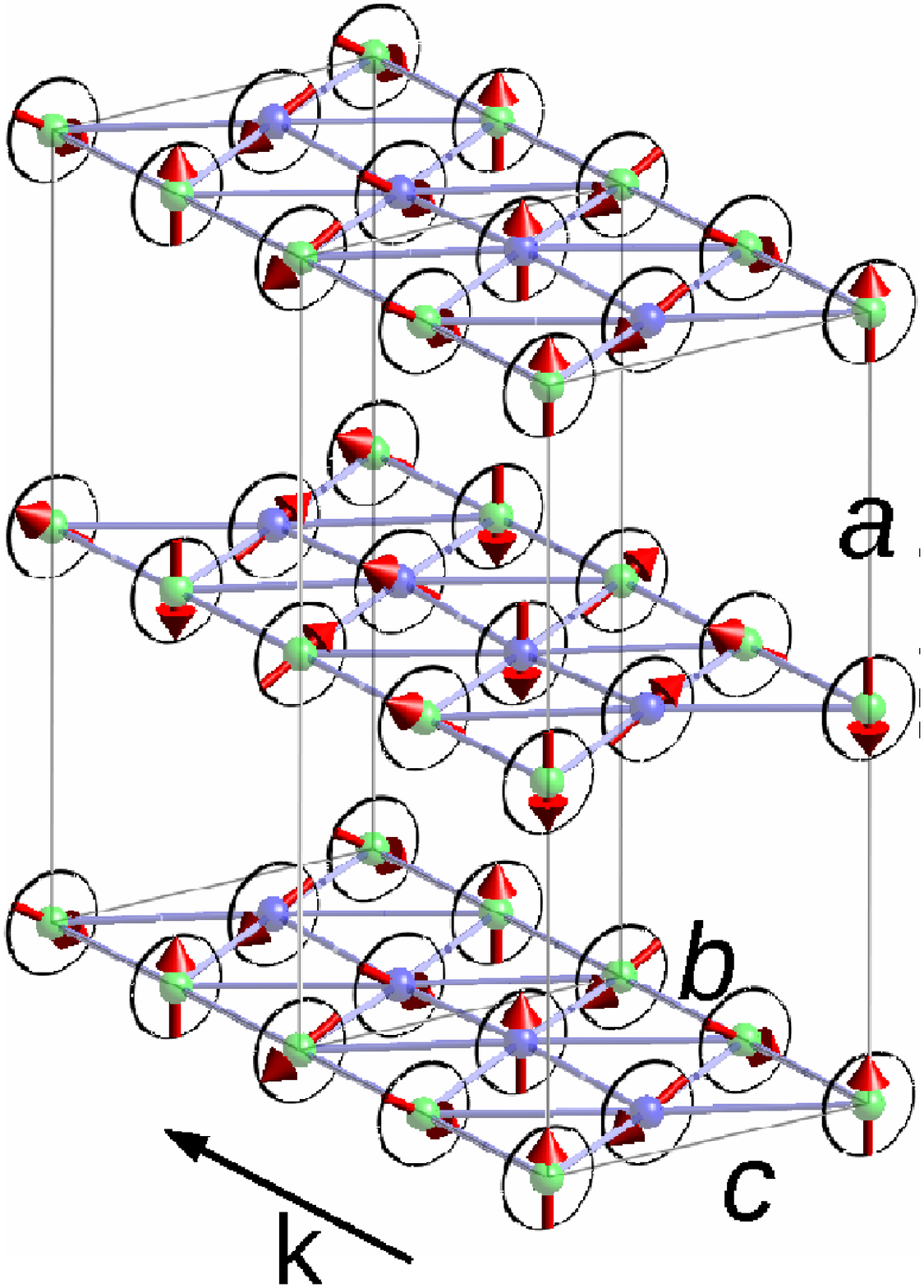}
\includegraphics[scale=0.37]{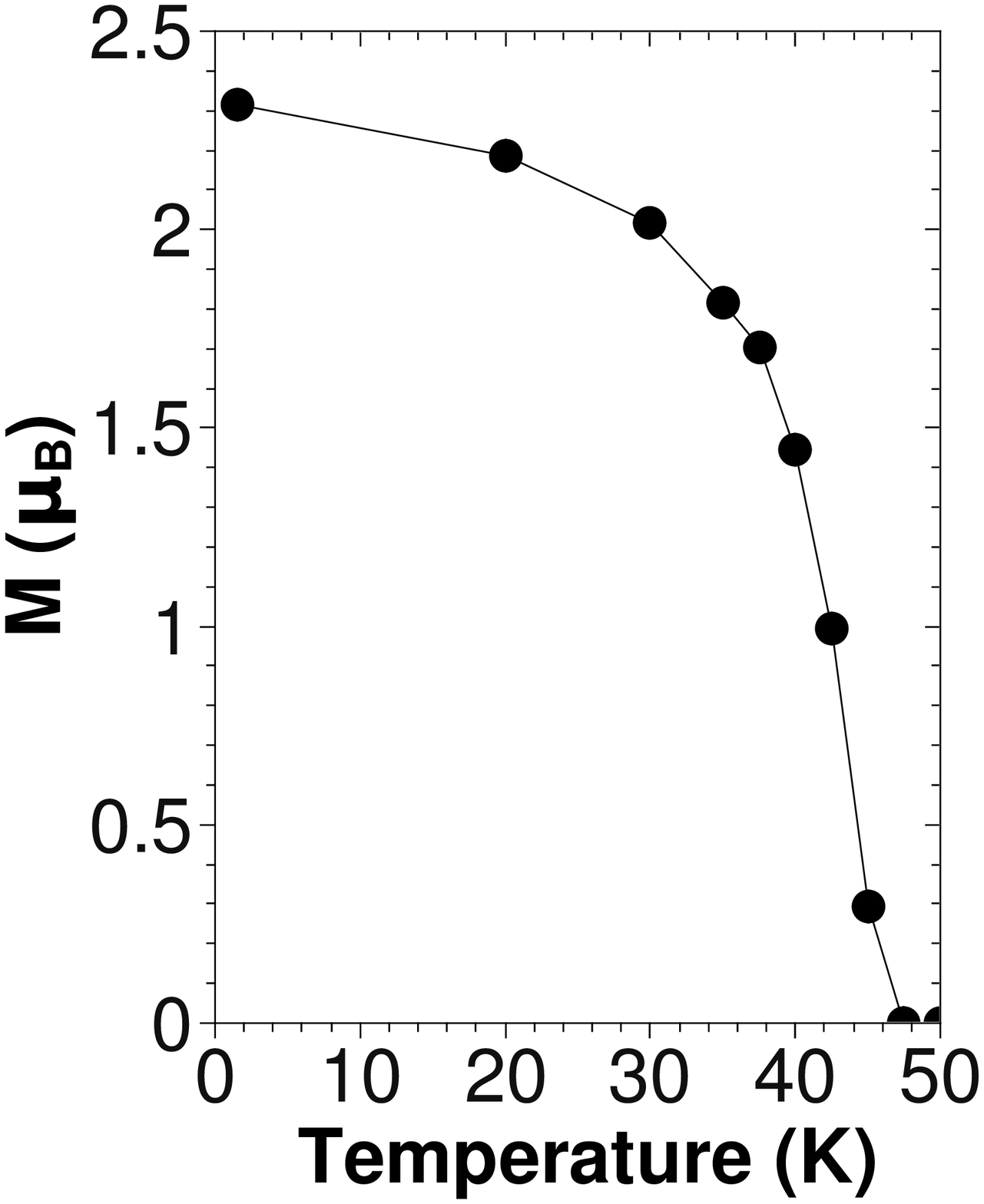}
\caption{(Color online) Left: Magnetic structure of \CACR\. The two independent Cr sites are shown with different colors (light green for Cr$_1$ and light blue for Cr$_2$). Two unit-cells along the \textit{b}-axis, marked by thin black lines, are shown. The axis are labelled in italic and the direction of the magnetic propagation vector (k) is shown by the arrow. Right: Temperature-dependence of the ordered magnetic moment on the Cr ions}
\label{fig:magneticstructure}
\end{figure}

\begin{figure}[h!]
\includegraphics[scale=0.32]{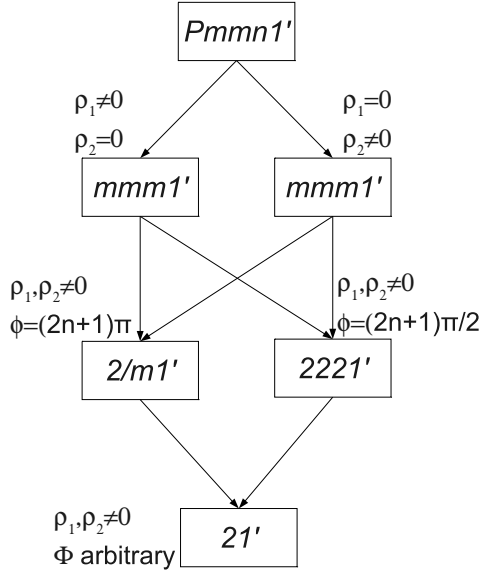}
\caption{Phase diagram of the possible magnetic point groups that can be stabilized from the 
paramagnetic group $Pmmn1'$ by considering two order parameters $\eta_1$ and $\eta_2$ transforming respectively as the $\tau_1$ and $\tau_2$ representations (see text for details).}
\label{fig:phasediagram}
\end{figure}

Symmetry analysis using the group of the propagation vector (little group) show that the magnetic representation for each Cr sites is decomposed in four irreducible representations, labelled $\tau_1$ to $\tau_4$ according to Kovalev's tables \cite{Kovalev}. The magnetic structure is only compatible with a model that mix a basis vector along the (1,0,0) direction that transforms as $\tau_1$ , and a basis vector along (0,0,1) that transform as $\tau_2$. In this orthorhombic symmetry, the intensity of the diffraction pattern is not sensitive to the relative phase between these two modes, i.e. one can not directly discriminate between a transversally polarized spin-density wave with the moments pointing in the \textit{ac}-plane (mixing the two basis vectors with coefficients of same characters) or a helical modulation with the moment rotating in the \textit{ac}-plane (mixing the two basis vectors with coefficients of different characters, real and imaginary). However at 1.5K, the refined amplitudes of the magnetic modes along the $a$ and $c$ axis were found to be 2.31(1) $\mu_B$. If these modes were in phase (spin-density wave) the amplitude of the modulation will largely exceed the fully ordered moment of 3 $\mu_B$ expected for a pure spin-state Cr$^{3+}$ ion. On this ground, it appears that the helical modulation is the only possible model to account for the observed pattern. It is also physically sensible as the entropy of an helical structure with constant moments is more favorable at low temperature than an amplitude modulated structure. The magnitudes of the magnetic moments were found to be similar for the two symmetry inequivalent Cr sites, and final refinements were conducted imposing such constraint, even if the latter is not directly imposed by symmetry. Finally, the relative phase between the modulation for site 1 and site 2 is a free parameter, as these sites are not related by symmetry operations of the group. The phase has been refined to a value very close to -$\frac{\pi}{3}$ and was constrained to this specific value in the final refinement. The magnetic structure, showed in Fig. \ref{fig:structure}, is made of nearly 120$^{\circ}$ configuration in each triangle, propagating as spirals along the \textit{b}-axis of the unit-cell with the two spirals per unit-cell (on different Cr sites) of same chirality. Unlike cycloidal modulations, this proper spiral structure possess a true handedness. The coupling between adjacent CrO$_2$ layers, mediated through super-superexchange interactions, is antiferromagnetic. The magnetic structure is similar to that found in delafossites \cite{PhysRevB.79.014412,PhysRevB.78.140401}, i.e. the 120$^{\circ}$ configuration is \emph{out-of-plane}. However, symmetry considerations lead to different coupling between magnetic order parameters and electric order than the delafossite, that will be discussed later. The ordered magnetic moment varies smoothly with temperature (Fig. \ref{fig:magneticstructure}) but the critical exponent was found to be $\sim\frac{1}{4}$, deviating largely from the mean-field limit and reminiscent to that found in first-order phase transitions \cite{Toledano}. Within the temperature resolution of our experiment there is a unique transition at T$_N$, in agreement with measurements of the magnetic susceptibility and specific heat.\\
The magnetic model consists of a mixing of two modes belonging to two different irreducible representations, i.e. the transition from the paramagnetic phase to the incommensurate magnetic phase can be fully described by considering two two-dimensional complex order parameters $\eta_1$=$\rho_1$e$^{i\phi_1}$,$\eta_1^*$=$\rho_1$e$^{-i\phi_1}$ and  $\eta_2$=$\rho_2$e$^{i\phi_2}$,$\eta_2^*$=$\rho_2$e$^{-i\phi_2}$. The matrix representatives of the complete irreducible co-representations of the group of $\textbf{k}$ are presented in Table \ref{Table:irrep}.
 The time-reversal operator 1$'$, which matrix is not shown, needs also to be considered to be able to write the Landau energy. Considering these two magnetic order parameters, the expansion of the Landau free energy F is written:\\
\begin{eqnarray}
F&=&F_0+\half \alpha_1 \rho_1^2 + \half \alpha_2 \rho_2^2 + \quarter \beta_1 \rho_1^4 + \quarter \beta_2 \rho_2^4 \\ \nonumber
&&+ \gamma \rho_1^2 \rho_2^2 cos(2 \phi) + ...\\
\label{eq:freeEnergy}
\end{eqnarray}
, where the $\alpha_i$, $\beta_i$ (i=1,2) and $\gamma$ are the usual coefficients and $\phi=\phi_1-\phi_2$ is the phase difference between the two modes. Minimization of F with respect to $\rho_1$ and $\rho_2$ leads to five distinct magnetic states that can be stabilized from the paramagnetic space group $Pmmn1'$. The point groups of these five magnetic states are shown in Fig. \ref{fig:phasediagram} together with their stability conditions. The structure observed experimentally corresponds to both order parameters being non-zero and $\phi=\pi/2$ (mixing of two modes in phase quadrature) and has the point symmetry 2221'. The 1' operator is contained in the point group of any incommensurate structure, since its application (phase of $\pi$) is always equivalent to a translation. The phase 2221' can be stabilized from the paramagnetic state either through two successive second-order phase transitions as indicated by the arrow in Fig. \ref{fig:phasediagram} or through a single first-order phase transition. According to our data showing a unique critical point but no pronounced hysteresis, we assumed that the transition must be \emph{weakly} first-order. One should note that the only possibility to stabilize the observed magnetic phase through a second-order transition would be to consider a symmetry for the exchange Hamiltonian that is higher than the crystal symmetry (for example considering an Heisenberg Hamiltonian). In such case, the symmetry analysis shows that the two irreducible representations involved, $\tau_1$ and $\tau_2$, belongs to the same exchange multiplet. However, this is rarely observed and always a case of pseudo-symmetry. In the following discussion, we will assume that the crystal symmetry is the relevant basis of work, and that the transition is weakly first-order.\\
\indent  The Landau theory of transitions with two order parameters that are coupled bi-quadratically has been studied by several authors \cite{Holakovsky1973,JPSJ.63.2082} and reviewed by Toledano et al. \cite{ToledanoReconstructive}. The invariant as written in Eq. \ref{eq:freeEnergy} is not sufficient to stabilize a first-order transition Pmmn1'$\rightarrow$2221' which would have to occur at a single point in the parameter space. In order to stabilize this transition through a line in phase space, one needs to consider a strongly negative coupling term $\gamma$ and at least one positive six degree term (for example $\delta\rho_1^6$). In such conditions, considering that the first coefficient $\alpha_1$ follows the usual temperature dependence $\alpha_1$=$\alpha(T-T_c)$ with $\alpha>0$, and that $\eta_1$ is the driving order parameter, the critical behavior of $\eta_2$ follows the triggering mechanism first discussed by Holakovsky \cite{Holakovsky}.\\ 
\indent The symmetry of this helical phase in such orthorhombic symmetry is extremely interesting. As already mentioned, the point group 2221', although not centrosymmetric, is not polar and therefore forbids a spontaneous polarization at the magnetic transition. Since the structure is incommensurate and the time-reversal operator is present in the point group, the linear magnetoelectric effect is also forbidden. However, invariants that are linear in the electric field (E) and quadratic in the magnetic field (H) are allowed by symmetry (whilst terms in HHE are forbiddden). The quadratic magnetoelectric effect is described by a third-rank tensor, $\beta_{ijk}$ symmetric in the last two indices and where i is the direction of the electric field and j and k indicate the direction of the magnetic fields. For the point group $2221'$ there are three terms allowed for this tensor that transform as the piezoelectric tensor, $\beta_{123}$,$\beta_{213}$ and $\beta_{312}$. One can easily show using the matrix representatives of $\Delta_2$ and $\Delta_3$ given in Tab. \ref{Table:irrep} that the following mixed invariants of the magnetic order parameters with E and H can be constructed: 
\begin{eqnarray}
F_{ME}=\rho_1 \rho_2 sin(\phi) \left[\beta_{123} E_xH_yH_z+\beta_{213} E_yH_xH_z
+\beta_{312} E_zH_xH_y \right]
\label{eq:ME}
\end{eqnarray}     
This suggests that the application of low magnetic fields (if the system remains in the phase described by the symmetry of the zero-field phase) inclined with respect to the crystallographic axes will induce an electrical polarization whose magnitude with be proportional to the square of the magnetic field. Of course, the application of sufficiently large magnetic fields could drive the system to other symmetry phases. Single crystal specimen are necessary to determine the behavior of the electric properties and magnetic properties under a magnetic field.\\
\indent It is important to realize that this result can be easily generalized to  all incommensurate magnetic systems that possess a true magnetic handedness, i.e. all spirals in the unit-cell propagate with the same chitrality. In such systems and irrespective of the paramagnetic symmetry, all improper symmetry operations are lost in the magnetically ordered state and the time-reversal operator is preserved by the incommensurability, implying the at least one of the $\beta_{ijk}$ is allowed and quadratic magnetoelectric effects are present. This magnetoelectric effect, even though of higher degree in the expansion of the free energy than the well-studied linear effect, is not bounded by thermodynamic conditions (see \cite{ISI:000239792700035} and references therein) and therefore interesting to investigate in prototype spiral magnets of the type discussed here.\\    
\indent In conclusion, measurements of the magnetic properties of \CACR\ and neutron powder diffraction have shown that this system is analog to the delafossite. Strong antiferromagnetic interactions in the triangular CrO$_2$ planes leads to long-range magnetic ordering below T$_N$=43K, with an incommensurate propagation vector k=(0,0.3317(2),0). The model for the magnetic structure, based on refinement of the neutron diffraction data and symmetry considerations, is consistent with an helicoidal modulation and a nearly 120$^{\circ}$ configuration in each triangular plaquette. The mirror plane symmetries are broken by the helicoidal arrangement, leaving the point group 2221$'$, which allows for quadratic magnetoelectric effects in all directions.\\     

\end{document}